\documentclass[conference,a4paper,10pt]{IEEEtran}
\usepackage{cite}
\usepackage[dvips]{graphicx}
\usepackage {times}
\usepackage{graphicx}
\usepackage{color}
\usepackage{amsmath}
\usepackage{amstext}
\usepackage{hyperref}
\usepackage{amsfonts}
\usepackage{amssymb}
\usepackage{psfrag}
\usepackage{fancyhdr}
\usepackage{dsfont}
\usepackage{bbm}
\usepackage{blkarray}
\usepackage{subfigure}
\usepackage{epsfig}
\usepackage[ruled]{algorithm}
\usepackage{algpseudocode}
\usepackage{tabularx}

\IEEEoverridecommandlockouts

\begin{document}

\title{Content-Aware User Clustering and Caching in Wireless Small Cell Networks}
\author{
\IEEEauthorblockN{Mohammed S. ElBamby\IEEEauthorrefmark{1}, Mehdi Bennis\IEEEauthorrefmark{1}, Walid Saad\IEEEauthorrefmark{2} and Matti Latva-aho\IEEEauthorrefmark{1} \\}
\IEEEauthorblockA{\small\IEEEauthorrefmark{1}Centre for Wireless Communications, University of Oulu, Finland, \\ email: \{melbamby,bennis,matti.latva-aho\}@ee.oulu.fi \\
\IEEEauthorrefmark{2}Wireless@VT, Bradley Department of Electrical and Computer Engineering, Virginia Tech, Blacksburg, VA, USA, email: walids@vt.edu}
\thanks{This research is supported by the SHARING project under Finland grant 128010 and the U.S. National Science Foundation (NSF) under Grants CNS-1253731 and CNS-1406947.}

\vspace{-.6cm}

}
\IEEEpubid{\makebox[\columnwidth]{978-1-4799-5863-4/14/\$31.00~\copyright~2014 IEEE \hfill}\hspace{\columnsep}\makebox[\columnwidth]{}}

\maketitle

\begin{abstract}
In this paper, the problem of content-aware user clustering and content caching in wireless small cell networks is studied. In particular, a service delay minimization problem is formulated, aiming at optimally caching contents at the small cell base stations (SCBSs). To solve the optimization problem, we decouple it into two interrelated subproblems. First, a clustering algorithm is proposed grouping users with similar content popularity to associate similar users to the same SCBS, when possible. Second, a reinforcement learning algorithm is proposed to enable each SCBS to learn the popularity distribution of contents requested by its group of users and optimize its caching strategy accordingly. Simulation results show that by correlating the different popularity patterns of different users, the proposed scheme is able to minimize the service delay by $42\%$ and $27\%$, while achieving a higher offloading gain of up to $280\%$ and $90\%$, respectively, compared to random caching and unclustered learning schemes.
\\ \\
\emph{Keywords- small cell networks; clustering; caching; offloading; reinforcement learning}

\end{abstract}

\IEEEpeerreviewmaketitle
\vspace{-.2cm}
\section{Introduction}
A tremendous increase in the demand for spectrum is expected over the next years, driven by the increasing need for mobile video streaming. Currently, $50\%$ of the mobile video traffic pertains to video streaming applications \cite{Cisco}. Such an increase in data traffic will require significant changes to today's cellular networks. One such change, the introduction of small cell base stations, is viewed as a key paradigm to handle the increase of video traffic and improve the wireless capacity by bringing contents closer to the users. However, reaping the benefits of small cell deployments requires meeting several key challenges such as resource allocation and network modeling \cite{guvenc_book}. Small cells also present new opportunities for network operators. For instance, owing to the cheap storage/memory prices and the fact that mobile video accounts for most of the total internet traffic, one can leverage the use of storage at the small cell level to bring popular contents closer to the network edge (i.e., BS and UE). Indeed, one promising approach to improve the quality-of-service (QoS) of video transmission is through caching popular contents locally at the small cell base stations to alleviate peak traffic demands and minimize service delays.

The use of caching in a backhaul-constrained small cell network is studied in \cite{femto_caching} using optimization algorithms. Leveraging device-to-device (D2D) communications for caching is studied in \cite{d2d_caching}. In \cite{5_5G}, the disruptive role of caching content in 5G cellular networks is discussed. The authors in \cite{mehdi_caching} study the benefits of both spatial and social caching as a means of enhanced traffic offloading in small cell networks. However, most of these existing works assume similar popularity patterns for all users in the system and do not consider the case in which users might have different interests over contents.

The main contribution of this paper is to study the problem of content-caching in wireless small cell networks. In particular, we consider a network in which small-cell users have different preferences over different content types. Consequently, there is a need to develop a novel scheme that allows to minimize the service delay by bringing popular contents close to the end users. To this end, we propose a spectral clustering approach, analogous to \cite{livehoods}, in order to group users into judiciously selected clusters based on the content similarity, which allows the small cell base stations (SCBSs) to effectively cache the most popular contents, and thus maximize the cache hit rates. Following the clustering phase, each cluster is associated to a different SCBS. By allowing each SCBS to service users that have similar content popularity distribution, the proposed caching policy enables SCBSs to prefetch users' popular contents to minimize the service delay. To dynamically update the SCBS caching strategy, a regret minimization learning algorithm is proposed allowing each SCBS to decide which content to cache. Simulation results show that by using the proposed scheme, the SCBSs are able to efficiently group users into clusters based on their content requests similarity. Given this clustering, SCBSs are able to adopt the proposed learning approach to minimize the delay for delivering users' content. Numerical results show that offloading gains can be achieved by caching more popular contents in the SCBS as compared to classical unclustered approaches.

The rest of this paper is organized as follows. Section \ref{sec:sys_form} introduces the system model. The proposed caching scheme is presented in Section \ref{sec:prop_algo}. In Section \ref{sec:sim_results}, we evaluate the performance of the proposed scheme, while conclusions are drawn in Section \ref{sec:conc}.

\vspace{-.2cm}
\section{System Model}
\label{sec:sys_form}
\begin{figure}
\centering
\includegraphics[width = 7.7cm, trim=0cm 0.6cm 0.5cm 0cm, clip=true]{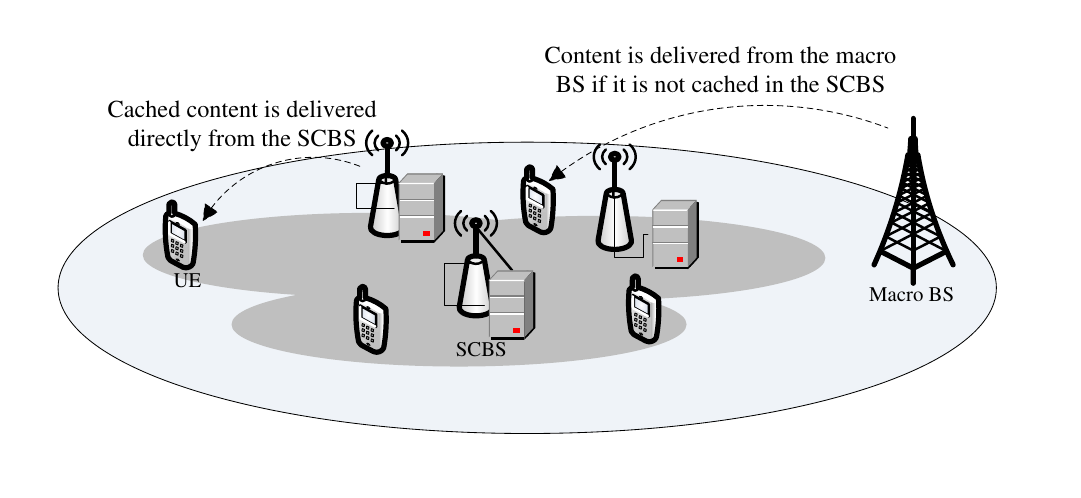}
\vspace{-.2cm}
\caption{Illustration of the considered network layout.}
\label{fig:layout}
\vspace{-.5cm}
\end{figure}

Consider a wireless heterogeneous network which consists of a macro base station (MBS) and a set $\mathcal{B}=\{1,\dotsc,B\}$ of SCBSs each serving a subset of user equipments (UEs). There are two types of UEs, macro UEs (MUEs), which are connected to the MBS, and small cell UEs (SUEs). Each SCBS $b \in \mathcal{B}$ is equipped with a data storage of capacity $f_b$ that contains $\mathcal{C}_b\subseteq\mathcal{C}$ from the available contents $\mathcal{C}$ in the system, i.e., a set of cached data. A SUE $u$ requests a certain content (file) $c \in \mathcal{C}$ with a request arrival rate that follows a Poisson distribution with mean $\lambda_{u,c}$ where a higher mean arrival rate reflects a higher content popularity. If a SUE requests a file that is available in the cache of its serving SCBS, it will be delivered directly, otherwise the MBS will handle the request, which can incur higher delay and higher costs. An illustration of the network layout is depicted in Fig. \ref{fig:layout}.

We assume that the content $c$ consists of a file of size $l_c$. We further define $D_{u,c}^{(b)}$ as the service delay experienced by UE $u$ to retrieve the requested content $c$ from BS $b$, expressed as follows:
\begin{equation}\label{eq:delay}
D_{u,c}^{(b)} = \frac{l_c}{r_{u,b}},
\vspace{-.2cm}
\end{equation}
where $r_{u,b}$ is the downlink (DL) transmission rate to user $u$ from base station $b\in\{0\}\cup\mathcal{B}$. Here, the index $0$ denotes the MBS, and $r_{u,b}$ is given by:
\begin{equation}\label{eq:rate}
r_{u,b} = w_{u,b}\log_2(1+\Gamma_{u,b}^{\text{DL}}),
\vspace{-.15cm}
\end{equation}
where $w_{u,b}$ is the bandwidth allocated to user $u$ being the total bandwidth for base station $b$ divided by the number of requests it serves, and $\Gamma_{u,b}^{\text{DL}}$ is the DL signal-to-interference-plus-noise-ratio (SINR).

Due to the storage capacity limits as well as the possible overhead for caching, it is not possible to cache all contents at all SCBSs. Therefore, there is a need for a cache replacement policy using which some of the cached contents are discarded while new contents are replaced. The main objective for each SCBS is to find the optimal caching strategy that minimizes the total delay $\sum_{u\in\mathcal{U}}\sum_{c\in\mathcal{C}}D_{u,c}^{(b)}$, where $\mathcal{U}$ is the set of SUEs. Here, a caching policy is defined as the probability distribution $\boldsymbol{\pi_b} = [\pi_{b,1},\dotsc,\pi_{b,C}]$ where each element $\pi_{b,k}$ represents the probability with which an SCBS will request content (or file) type $k$. We assume that there is a total of $C$ file types.

Given that the popularity of a certain content can vary between different UEs, there is a need to cluster the UEs based on their content request similarities. In other words, each SCBS should serve a group of users with similar interests in certain contents so as to optimize its caching strategy accordingly. Consequently, we define $\boldsymbol{Q}$ as the vector of UE-SCBS associations such that $q_u\in \boldsymbol{Q}$ represents the SCBS $b$ which is servicing UE $u$. Consequently, the joint clustering and caching optimization problem is formulated as follows:
\vspace{-.1cm}
\begin{align}\label{eq:opt_prob}
& \underset{\boldsymbol{Q},\boldsymbol{\pi_b}}{\text{minimize}}
& &J(\boldsymbol{Q},\boldsymbol{\pi_b}) = \sum_{b\in\mathcal{B}}\sum_{u:q_u=b}\sum_{c\in\mathcal{C}}D_{u,c}^{(b)},
\end{align}
\vspace{-.55cm}
\begin{align}
& \text{subject to}
& & 0\leq \pi_{b,c}\leq1, \; \forall c \in \mathcal{C}, \forall b \in \mathcal{B}, \label{const1}\\
&
& &  q_u\in  [1,\dotsc,B], \; \forall u \in \mathcal{U}, \label{const2}\\
&
& & \sum_{c\in \mathcal{C}_b}l_c\leq f_b, \; \forall b \in \mathcal{B}, \label{const3}
\end{align}
where \eqref{const1} is the probability constraint, \eqref{const2} is the UE association constraint and \eqref{const3} is the storage capacity constraint of SCBS $b$.
\vspace{-.2cm}
\section{Regret-Based Caching Scheme}\label{sec:prop_algo}
In this section, we propose a joint clustering and caching scheme to solve the optimization problem described in \eqref{eq:opt_prob}. To solve the problem in a decentralized manner, we decouple the problem into two subproblems. First, we group UEs into clusters based on their file requests during a training period. Then, each cluster of UEs is associated with a specific SCBS where the caching learning procedure is done locally.
\subsection{Content-Based User Clustering}
Contrary to traditional location-based proximity clustering methods, we use the notion of \emph{content proximity} to develop a clustering algorithm that groups users based on their content-based similarities. In particular, we use a spectral clustering \cite{spectral_clustering} algorithm to discover similarities between users requesting similar contents. Here, we consider a training period during which UEs submit their content requests to the network. Then, we build a similarity matrix between UEs either at the level of the MBS or SCBSs, assuming that the SCBSs communicate during the training period. After a training period of $t_t$ time instants, for each UE $u$, we build a content frequency vector $\boldsymbol{n}_u=[n_{u,1},\dotsc,n_{u,C}]$ where $n_{u,c}$ means that content $c$ was requested by user $u$, $n_{u,c}$ times. This frequency of requests can be seen as an approximation to the content mean arrival rate $\lambda_{u,c}$ during the training period. Subsequently, we use the cosine similarity metric to measure the similarity between users $i$ and $j$ requesting similar contents as follows:
\begin{equation}\label{eq:similarity}
s(i,j) = \frac{\boldsymbol{n}_i.\boldsymbol{n}_j}{\|\boldsymbol{n}_i\|\|\boldsymbol{n}_j\| }
\end{equation}
Algorithm \ref{algo:clustering_algorithm} describes the proposed UE clustering method. After the clustering, each group of users is associated to a specific SCBS. It is worth noting that with this user association, a user might not be associated to the SCBS with higher SINR. However, by grouping similar users together, more popular content will be cached closer to users. This results in less requests being served from the macrocell, minimizing the total service delay.

\begin{algorithm}[t] 
 \begin{algorithmic}[1]
 \State \textbf{Initialization}: pick a sequence of time instants, calculate the vector of requests occurrence $\boldsymbol{n}_u$ for each user, calculate the similarity matrix $\boldsymbol{S}=[s(i,j)]$ as in (\ref{eq:similarity}), choose $k_{\text{min}}=2$ and $k_{\text{max}}=U/2$.
 \State Calculate the diagonal degree matrix $\boldsymbol{D}$ with diagonal element $d_i=\sum_{j=1}^{U}s_{i,j}$.
 \State Calculate $\boldsymbol{L}=\boldsymbol{D}-\boldsymbol{G}$.
 \State Calculate $\boldsymbol{L}_{\text{norm}}=\boldsymbol{D}^{-1/2}\boldsymbol{L}\boldsymbol{D}^{-1/2}$.
 \State Pick a number of $k_{\text{max}}$ eigenvalues of $\boldsymbol{L}_{\text{norm}}$ such that $\lambda_1 \leq \dotsb \leq \lambda_{k_{\text{max}}}$.
 \State Choose $K= \max_{i=k_{\text{min},\dotsc,k_{\text{max}}}}\Delta_i$ where $\Delta_i =\lambda_{i+1}-\lambda{i}$.
 \State Calculate the $k$ smallest eigenvectors $x_1,\dotsc,x_k$
 \State Let the $\boldsymbol{Y}$ matrix has the eigenvectors $x_1,\dotsc,x_k$ as columns.
 \State Use the k-means clustering to cluster the rows of the matrix $\boldsymbol{Y}$.
 \end{algorithmic}
 \caption{Clustering Algorithm}
 \label{algo:clustering_algorithm}
 \vspace{-.1cm}
\end{algorithm}

\subsection{Distributed Caching Strategy}
Given the UE clustering, our next step is to propose a decentralized caching scheme to minimize the delay incurred to deliver requests to users, where the utility of SCBS $b$ is $\bold{v}_b(\boldsymbol{\pi_b})=1/J_b(\boldsymbol{\pi_b})$ where $J_b(\boldsymbol{\pi_b}) = \sum_{u:q_u=b}\sum_{c\in\mathcal{C}}D_{u,c}^{(b)}$. Here, each SCBS is interested in following a caching policy that minimizes the service delay for its associated group of UEs. The proposed caching scheme is decentralized and is performed in each SCBS using local information. In particular, SCBSs are able to learn the probability distribution of their caching strategies by minimizing their \emph{regret} over caching the content in the past and using this information to optimize the caching decisions in the next time instants.

The proposed scheme is based on the distributed regret learning approach inspired from \cite{mehdi_learning}. That is at each time instant $t$, each SCBS $b$ picks up an action $a_b^{(n_b)}$, that is a binary value that determines whether to cache the content $c$ or not, from the action space $ \mathcal{A}_b = \{a_b^{(1)},a_b^{(2)}, \dotsc, a_b^{(N_b)}\}$ where $N_b$ is the total number of actions which is equal to the total number of available content. This content is then cached replacing another existing content in the storage. Each SCBS chooses an action $a_b^{(n_b)}$ following the probability distribution of all actions $\boldsymbol{\pi}_{b}(t)=[\pi_{b,a_b^{(1)}}(t),\pi_{b,a_b^{(2)}}(t),\dotsc,\pi_{b,a_b^{(N_b)}}(t)]$, where $\pi_{b,a_b^{(n_b)}}(t)$ is the probability that SCBS $b$ plays the action $a_b^{(n_b)}$ at time instant $t$, i.e.,
\begin{equation}
\pi_{b,a_b^{(n_b)}}(t)= \Pr(a_b(t)=a_b^{(n_b)}).
\end{equation}
In the proposed approach, each SCBS keeps updating its actions following a specific strategy vector $\boldsymbol{\pi}_{b}$, which is the probability distribution of choosing an action. Then, it compares its time-average observation of the utility function with the case in which it plays this same action in all previous time instants. In this regard, each SCBS will be interested in choosing a probability distribution that minimizes its regret of playing/not playing each action. This balances the tradeoff between caching the most popular contents and having a non-zero probability of caching the less popular files. Therefore, each SCBS $b$ estimates its utility vector $\hat{\bold{v}}_b(t)=[\hat{v}_{b,a_b^{(1)}}(t),\dotsc,\hat{v}_{b,a_b^{(N_b)}}(t)]$ and regret vector $\boldsymbol{\hat{r}}_b(t)=[\hat{r}_{b,a_b^{(1)}}(t),\dotsc,\hat{r}_{b,a_b^{(N_b)}}(t)]$ for each action assuming it has played the same action during all previous time instants $\{1,\dotsc,t-1 \}$.
In this scheme, SCBSs aim at minimizing their regret while estimating their time-average utility from playing a particular action at time instant $t$. The objective is to minimize the regret of caching or not caching a specific file, e.g. a file is cached but not requested. Hence, a non-zero probability of caching a less popular file is needed. To balance this tradeoff, each SCBS will choose the actions that yield higher regrets more likely than those yielding lower regrets, keeping a non-zero probability for playing any actions. This behavior is captured by the Gibbs Sampling-based probability distribution, in which the probability of playing an action $a_b^{(n_b)}$ by an SCBS $b$ can be expressed as follows:
\begin{equation}\label{eq:gibbs}
\Lambda_{b,a_b^{(n_b)}} (\boldsymbol{a}_{-b}) = \frac{\exp\left(\beta_b r_b^{+}(a_b^{(n_b)},\boldsymbol{a}_{-b})\right)} {\sum_{m=1}^{N_b}\exp\left(\beta_b r_b^{+}(a_b^{(m)},\boldsymbol{a}_{-b})\right)},
\end{equation}
where $\beta_b$ is a Boltzmann temperature coefficient that controls the above-mentioned tradeoff, and $\boldsymbol{r}_b^{+}(t)$ denotes the vector of positive regrets $\boldsymbol{r}_b^{+}(t)=\max(0,\boldsymbol{r}_b(t))$.However, maximizing the SCBS utility function (i.e., minimizing the service delay) depends not only on its own choice of action but also on remaining BSs due to the interference and the throughput from the macrocell. Therefore, at each time instant $t$, a SCBS $b\in\mathcal{B}$ estimates $\hat{\bold{v}}_b(t)$,$\boldsymbol{\hat{r}}_b(t)$ and $\boldsymbol{\pi}_b(t)$ using a regret learning process as follows:
\begin{align}\label{eq:cost_and_strategy_learn}
\begin{cases}
\hat{v}_{b,a_b^{(n_b)}}(t) = \hat{v}_{b,a_b^{(n_b)}}(t-1) + \\
 \hphantom{1cm}\alpha_b(t).\mathds{1}_{\{a_b(t-1)=a_b^{(n_b)}\}}\left(\tilde{v}(t-1) - \hat{v}_{b,a_b^{(n_b)}}(t-1) \right),\\
 \hat{r}_{b,a_b^{(n_b)}}(t) = \hat{r}_{b,a_b^{(n_b)}}(t-1) + \\
 \hphantom{1cm}\gamma_b(t).\left(\hat{v}_{b,a_b^{(n_b)}}(t-1)-\tilde{v}(t-1) - \hat{r}_{b,a_b^{(n_b)}}(t-1) \right),\\
\pi_{b,a_b^{(n_b)}}(t) = \pi_{b,a_b^{(n_b)}}(t-1) +\\
\hphantom{10}\hphantom{10}\hphantom{10}\hphantom{10} \zeta_b(t).\left(\Lambda_{b,a_b^{(n_b)}} \bigl( \boldsymbol{\hat{r}}_b(t-1) \bigr) - \pi_{b,a_b^{(n_b)}}(t-1) \right),
\end{cases}
\end{align}
where $\tilde{v}(t-1)$ is the instantaneous observed utility function at time $t-1$, $\Lambda_{b,a_b^{(n_b)}}$ is given by (\ref{eq:gibbs}), $\alpha_b(t)$, $\gamma_b(t)$ and $\zeta_b(t)$ are the learning parameters, and should satisfy the following constraints \cite{mehdi_learning}:
\begin{align}\label{eq:converging_conditions}
\vspace{-.6cm}
\begin{cases}
   (i)\lim\limits_{T\rightarrow\infty}\sum\limits_{t=1}^{T}\alpha_b(t)=+\infty,\lim\limits_{T\rightarrow\infty}\sum\limits_{t=1}^{T}\alpha_b(t)^2<+\infty,\\
   (ii)\lim\limits_{T\rightarrow\infty}\sum\limits_{t=1}^{T}\gamma_b(t)=+\infty,\lim\limits_{T\rightarrow\infty}\sum\limits_{t=1}^{T}\gamma_b(t)^2<+\infty,\\
   (iii)\lim\limits_{T\rightarrow\infty}\sum\limits_{t=1}^{T}\zeta_b(t)=+\infty,\lim\limits_{T\rightarrow\infty}\sum\limits_{t=1}^{T}\zeta_b(t)^2<+\infty,\\
   (iv)\lim\limits_{t\rightarrow\infty}\frac{\zeta_b(t)}{\gamma_b(t)}=0, \lim\limits_{t\rightarrow\infty}\frac{\gamma_b(t)}{\alpha_b(t)}=0.
\end{cases}
\vspace{-.6cm}
\end{align}
This process guarantees the convergence of the algorithm to an \emph{$\epsilon$-coarse correlated equilibrium} \cite{mehdi_learning}.
\subsection{Cache removal scheme}
We consider a cache removal mechanism to select which existing file should be replaced by the cached file. At each time instant $t$, an SCBS chooses to cache a new content that is not in its storage. If the storage of the SCBS is already full, then it has to remove one of the existing contents. To be able to remove an appropriate content, each SCBS builds a content popularity vector based on the frequencies of requests $\boldsymbol{n}_b=[n_{b,1},\dotsc,n_{b,C}]$. Consequently, contents with lower frequencies of being requested are a better candidate for being removed. The Gibbs-Sampling probability distribution is used to remove the content from the cache:
\begin{equation}\label{eq:gibbs2}
G_{b,c} (\boldsymbol{n}_b) = \frac{\exp\left(-\beta_{\text{remove}}. n_{b,c}\right)} {\sum_{m=1}^{C}\exp\left(-\beta_{\text{remove}}. n_{b,m}\right)}
\end{equation}
where $\beta_{\text{remove}}$ is the Boltzmann's temperature coefficient, and the negative sign is to give higher probabilities to the contents having a lower frequency of being requested. The use of the Gibbs-Sampling probability distribution allows to use the $\beta_{\text{remove}}$ parameter to update the cache removal policy. While using $\beta_{\text{remove}}=0$ gives all contents equal probability to be removed, higher values of $\beta_{\text{remove}}$ means that contents with lower request frequencies will be removed with higher probabilities.\\
\vspace{-.4cm}
\section{Simulation Results} \label{sec:sim_results}
In this section, we analyze the performance of the proposed content caching scheme. We assume that the popularity of different contents (files) in the system follows a Zipf distribution \cite{zipf_paper}. Following a Zipf popularity model, the popularity of content $i$ is given by:
\vspace{-.2cm}
\begin{equation}\label{eq:zipf}
\lambda_i=\frac{i^{-\alpha_z}}{\sum_{j\leq C}j^{-\alpha_z}}\bar{\lambda}
\vspace{-.1cm}
\end{equation}
where $\bar{\lambda}$ is the average content popularity and $\alpha_z$ is the Zipf parameter. This means that the request rate for the $i^{th}$ most popular content is proportional to $1/i^{\alpha_z}$. To be able to assess the performance of the UE clustering scheme, we assume that we have three types of UEs in the system. For each type, the order of content popularity is different, i.e., UEs have different preference over different contents. We are interested in the performance of the SUEs, for different values of the Zipf parameter $\alpha_z$. We assume that the macro cell divides the bandwidth equally between the requests of the MUEs and the SUEs that request content from the MBS if not cached in the SCBS. For the path loss model, we use the 3GPP baseline parameters \cite{standard_pathloss}. We compare the proposed scheme against two baseline schemes. The first one is a random caching scheme in which at each time instant, a SCBS picks a random content to cache, and if the storage is full, it removes a random content chosen uniformly. UEs are associated to the SCBS with the higher received signal strength indicator (RSSI). The second baseline scheme is based on the proposed regret learning caching scheme but without clustering. Simulation parameters are summarized in Table \ref{tab:parameters}.

\begin{table}[!t]
\renewcommand{\arraystretch}{1.3}
\caption{Simulation parameters}
\vspace{-.2cm}
\label{tab:parameters}
\centering
\begin{tabular}{p{1.7in}||p{1.1in}}
\hline
\bfseries Parameter & \bfseries Value/description\\
\hline\hline
Number of macro cells & 1\\
\hline
Number of SCBSs & 3\\
\hline
System bandwidth & 5 MHz\\
\hline
Small cell radius & 40 m\\
\hline
Number of MUEs & 50\\
\hline
Number of SUEs & 15\\
\hline
MBS transmission power & 46 dBm\\
\hline
SCBS transmission power & 30 dBm\\
\hline
Thermal noise & -174 dBm/Hz\\
\hline
Number of contents & 30\\
\hline
Average content popularity ($\bar{\lambda}$) & $10$ \\
\hline
\end{tabular}
\begin{tabular}{p{3.2in}}
\centering \textbf{Learning parameters}
\end{tabular}
\begin{tabular}{p{1.7in}||p{1.1in}}
\hline
Strategy learning rate ($\zeta_b)$&$1/(t)^{0.7}$\\
\hline
Regret learning rate ($\gamma_b)$&$1/(t)^{0.6}$\\
\hline
Utility learning rate ($\alpha_b)$&$1/(t)^{0.5}$\\
\hline
Regret temperature coefficient ($\beta_b$) &$20$\\
\hline
Cache removal coefficient ($\beta_{\text{remove}}$) &$10/t$\\
\hline
\end{tabular}
\vspace{-.2cm}
\end{table}

\begin{figure}
\centering
\includegraphics[width = 7.7cm, trim=3cm 8.5cm 3cm 9cm, clip=true]{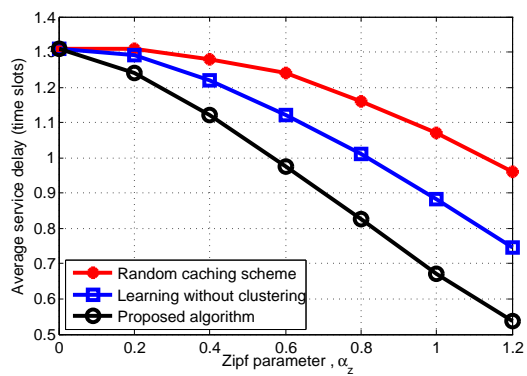}
\vspace{-.2cm}
\caption{Average service delay for different values of the Zipf parameter, $3$~SCBSs and a storage capacity of $10$.}
\label{fig:delay_zipf}
\vspace{-.4cm}
\end{figure}

\begin{figure}
\centering
\vspace{-.6cm}
\includegraphics[width = 7.7cm, trim=3cm 8.5cm 3cm 9cm, clip=true]{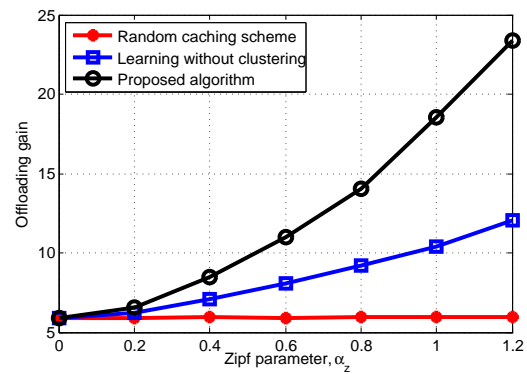}
\vspace{-.2cm}
\caption{Offloading gain for different values of the Zipf parameter, $3$~SCBSs and a storage capacity of $10$.}
\label{fig:offloading_zipf}
\vspace{-0.8cm}
\end{figure}

In the beginning of each simulation, we allow a training period of $500$ time instants ($0.5$ seconds) for carrying out the clustering procedure. In the beginning, we assume that the SUEs are associated to the SCBSs with the highest RSSI. The clustering algorithm is able to accurately group users into different clusters based on the received requests. Hence, users are grouped into clusters and each cluster is associated to a different SCBS.

We evaluate the performance of the proposed caching scheme. In Fig. \ref{fig:delay_zipf}, we show the average service delay for the proposed scheme against the baseline schemes for different values of the Zipf parameter $\alpha_z$. In this figure, we can see that the proposed algorithm achieves significant gains in terms of lower service delay (i.e., $42\%$ and $27\%$) as compared to random caching and learning without clustering schemes, respectively. For all schemes, the delay decreases as the Zipf parameter increases. This is due to the fact that having higher Zipf parameter with the same mean popularity $\bar{\lambda}$ leads to lower average request rates, which can be inferred from (\ref{eq:zipf}). The results also show that the gain of the proposed scheme over the random scheme increases as the Zipf parameter increases. This is because the random caching treats all files equally, so having files with higher popularity variance decreases the probability of serving these files from the small cells, hence degrading the performance. For the learning without clustering scheme, since users are not clustered, an SCBS has users with different interests for contents. Therefore, it is unable to learn the optimal strategy from few observations (i.e., cold-start problem). Instead, since the proposed scheme is able to cluster users with similar interest in one SCBS, the SCBS will be able to more efficiently learn the popular contents of its users and hence cache the most popular files accordingly.

Fig. \ref{fig:offloading_zipf} shows the offloading gain for different caching schemes, defined as the ratio between the SCBS throughput to the throughput obtained from the MBS. The results show that the proposed scheme achieves significant offloading gain by caching popular contents in the SCBSs. The gain increases as the Zipf parameter increases, since the file popularities become more diverse. Fig. \ref{fig:offloading_zipf} shows that this performance advantage reaches up to $280\%$ and $90\%$ relative to the random caching and learning without clustering schemes, respectively.

\begin{figure}
\centering
\includegraphics[width = 7.7cm, trim=3cm 8.5cm 3.5cm 9cm, clip=true]{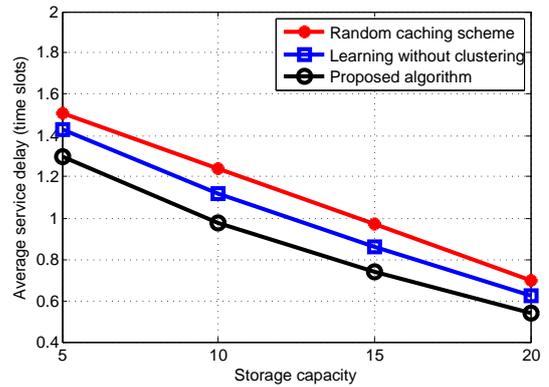}
\vspace{-.4cm}
\caption{Average service delay for different storage capacities, with a Zipf parameter $\lambda_z$ of $0.6$ and $3$~SCBSs.}
\label{fig:delay_storage}
\vspace{-.4cm}
\end{figure}

\begin{figure}
\centering
\includegraphics[width = 7.7cm, trim=3cm 8.5cm 3.5cm 9cm, clip=true]{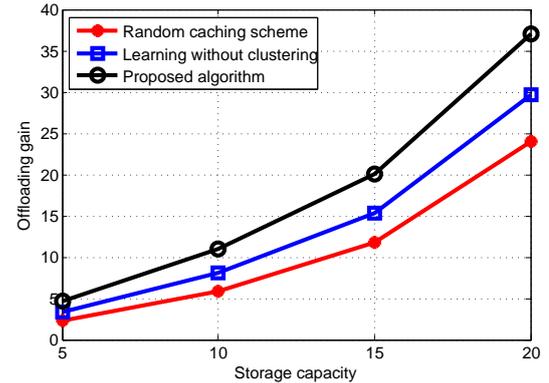}
\vspace{-.4cm}
\caption{Offloading gain for different storage capacities, with a Zipf parameter $\lambda_z$ of $0.6$ and $3$~SCBSs.}
\label{fig:offloading_storage}
\vspace{-.6cm}
\end{figure}

In Fig. \ref{fig:delay_storage} and Fig. \ref{fig:offloading_storage}, we show the average service delay and offload gain as the storage capacity varies. These figures show that, for all schemes, the service delay decreases and the offloading gain increases with increasing the storage size. This is due to the fact that the base stations will be able to provide more contents close to the end users. Moreover, the proposed scheme achieves much lower service delay and higher offloading gain compared to the baseline schemes. The gain increases as the storage capacity increases, since higher capacity allows the SCBSs to cache more popular contents following their caching strategies. Fig. \ref{fig:delay_storage} and \ref{fig:offloading_storage} show that this performance advantage reaches up to $24\%$ and $15\%$ lower delay and $65\%$ and $33\%$ higher offloading gain relative to the random caching and learning without clustering schemes, respectively.\\

\vspace{-.5cm}
\section{Conclusions}\label{sec:conc}
In this paper, we have proposed a joint user clustering and caching scheme for wireless small cell networks. The proposed approach allows to exploit the social similarities to group users into different clusters. Each cluster is then associated with a suitable SCBS. In this way, SCBSs are able to effectively cache the most popular contents and reduce service delays. Simulation results show that by bringing the popular contents close to the small cell UEs, the proposed algorithm outperforms random caching and learning without clustering schemes in terms of lower service delay and higher offloading gain.






\end{document}